\documentstyle[12pt]{article}
\begin{document}
\begin{center}
{\Large \bf
{ANTIGRAVITING BUBBLES WITH THE}\\
\vskip 0.3cm
{NON-MINKOWSKIAN ASYMPTOTICS}
} \\
\vskip 0.5cm
{ Andro BARNAVELI and Merab GOGBERASHVILI } \\
\vskip 0.5cm
{\it
 {Institute of Physics of the Georgian Academy of Sciences,}\\
 {Tamarashvili str. 6, Tbilisi 380077, Republic of Georgia.}\\
 {(E-mail: barbi@iberiapac.ge ; gogber@physics.iberiapac.ge ).}
} \\
\vskip 1cm
{\Large \bf Abstract}\\
\vskip 0.3cm
\quotation
{\small
The conventional approach describes the spherical domain walls by the
same state equation as the flat ones. In such case they also must be
gravitationally repulsive, what is seemingly in contradiction with
Birkhoff's theorem. However this theorem is not valid for the
solutions which do not display Minkowski geometry in the infinity.

In this paper the solution of Einstein equations describing the
stable gravitationally repulsive spherical domain wall is considered
within the thin-wall formalism for the case of the non-Minkowskian
asymptotics.
}
\endquotation
\end{center}

\vskip 1cm
For the last two decades great attention has been paid to the
investigation of gravitational properties of topological structures
such as domain walls, strings and monopoles. It was obtained, that
cosmic strings do not produce any gravitational force on the
surrounding matter locally while global monopoles, global strings
and planar domain walls exhibited repulsive
nature \cite{VS95,HL90,HS88}.  In this paper we shall consider some
problems which arose at studying the gravitational properties of
spherical domain walls and show the existence of the solution of
Einstein's equations corresponding to a stable gravitationally
repulsive spherical domain wall.

It is assumed, that the flat domain walls are described by the
 state equation \cite{VS95}:
\begin{equation}
\sigma  = - p = const,
\label{1}
\end{equation}
where $\sigma$ is the surface density and $p$ is the strong tension
in two spatial directions. This state equation corresponds to de
Sitter's expansion in the wall-plane and the borders of the wall
running away with the horizon. One can speak about the gravitational
field of the wall only in the normal direction to the wall. If, for
such objects, it is possible to use Newtonian approximation with the
mass described by Tolman's formula
\begin{equation} M = \int (T_0^0 -
 T_1^1 - T_2^2 - T_3^3) \cdot \sqrt{-g}dV = \int (\sigma+2p) \cdot
 \sqrt{-g}dV = - \int \sigma \cdot \sqrt {-g}dV \label{2},
\end{equation}
then the tension $p$ acts as a repulsive source of gravity and the
planar domain wall has a negative gravitational mass exhibiting
repulsive gravitational field \cite{VS95}.

It is natural to think that the same  behavior  (gravitational repulsion)
must occur for the  spherical  domain  walls  (bubbles), since
usually it is assumed that they are described by the same state
equations (\ref{1}) (e.g. see \cite{BKT87,IS84}).  On the other hand,
according to Birkhoff's theorem, the empty space, surrounding any
spherical body (including bubbles),  is described  by Schwarzschild
metric.  This metric contains the parameter $m$ (corresponding to the
mass of gravitating body)
\begin{equation}
 m = \int T_0^0 \cdot \sqrt{-g}dV ,
\label{3}
\end{equation}
which independently of the state equation is positive.  While for
planar domain walls (stretching the horizon) the negative
gravitational mass (\ref{2}) can be admissible, for bubbles the
negativeness of mass (\ref{3}) from the first glance looks
surprising, since $T_0^0$ is positively defined everywhere.

The above-mentioned problem emerged also when investigating bubble
dynamics within the thin-wall formalism \cite{I66}. It was obtained
that active gravitational mass of the spherical domain wall is
positive, i.e. its gravitational field is attractive
\cite{BKT87,IS84,L84}.  The disagreements in gravitational properties
of planar and spherical domain walls were explained by instability of
the latter \cite{IS84}, or by existence of a positive energy source
stabilizing the bubble \cite{L84}. However there still remain
various paradoxes (appearing in the models with large pressure
\cite{IS84,G85,BGG87}) which can be solved only if bubbles with
the state equation (\ref{1}) are repulsive.

The negative mass problem can be solved by the assumption that
domain walls are not described by the state equation (\ref{1}).
One must take into account the flux out from the volume of
integration, or some external forces stabilizing the domain wall. As a
result a state equation can have a principally different form and
both the spherical and planar walls can be gravitationally
attractive. The other possible solution of discrepancy may be the
assumption that the planar domain walls are described by the state
equation (\ref{1}) while the bubbles are not.

Recently we have investigated the bubble dynamics within the thin-wall
formalism when the state equation for spherical domain walls
nevertheless has the form (\ref{1}).  We have found a solution
describing repulsive spherical domain walls with outer
the Schwarzschild geometry \cite{BG94}, but only in the case when the
time coordinate changes its direction on the wall-surface.

In this paper we consider the different case, when the domain walls
are described by the state equation (\ref{1}), the time-flow has the
same direction in whole space, however the metric far from the
spherical domain wall is not Minkowskian. We show that in such case
there also exists a solution of the Einstein equations which
corresponds to a gravitationally repulsive stable bubble.

The assumption about non-Minkowskian asymptotics is reasonable, since
in the case of spherical domain walls it is impossible to  surround
the full source by any boundary inside the horizon (just as it is for
planar domain walls).  The domain wall is only the "part" of the
scalar field solution which fills the whole Universe up to horizon
and which has a nonzero vacuum expectation value even in the
infinity.  The result is that the quantity $\int T_{\mu\nu}\cdot
dS^{\nu} $ is not a 4-vector of energy-momentum and one can not
define the energy simply as $\int T_{00}\cdot dxdydz$.  For example,
the energy density of an expanding spherical domain wall remains
constant (see (\ref{1})) despite increasing of its surface, i.e.,
this object "takes" the energy from vacuum.

In pure Einstein's theory it has been proved that the total energy
carried by an isolated system, generating an asymptotic Minkowski
geometry, is positive \cite{SY}. Due to the essential role played by
the asymptotic condition this theorem can not be applied to solutions
of Einstein's theory which do not display a Minkowskian asymptotic
structure. In order to demonstrate that the sign of the gravitational
potential depends on the asymptotical geometry let us consider the
zero-zero component of the metric tensor for the isolated source in
Newton's approximation $$g_{00} = g_{00}^{\infty} + \Phi ,$$ where
$g_{00}^{\infty}$ is the asymptotic value of metric tensor and
\begin{equation} \Phi = g_{00} - g_{00}^{\infty} \label{4}
\end{equation}
is Newton's potential.

When far from the source we have Minkowskian geometry, then
$g_{00}^{\infty}$ reaches the maximal value, $1$, and, since $g_{00}
\le 1$, $\Phi$ is always negative, i.e.,  we have gravitational
attraction. For non-Minkowskian asymptotics, when $ g_{00}^{\infty} <
1 ,$ Newton's potential (\ref{4}) can be positive or zero depending
on the state equation of the source. The examples of sources with
non-Minkowskian asymptotics and with unusual gravitational behavior,
as it was mentioned above, are topological objects
\cite{VS95,HL90,HS88}.

Since the exact solution  of the coupling Einstein-Higgs equations
for the spherically domain wall is unknown we shall work within the
thin-wall formalism. Then Einstein's equations describing motion of
spherical domain walls in the case when the time-flow has the same
direction in whole space have the form \cite{BKT87,I66}:
\begin{equation}
\sqrt{f_+ + \dot R^2} - \sqrt{f_- + \dot R^2} = - \kappa GR ,
\label{5}
\end{equation}
where $\kappa = 4\pi \sigma$ and $f_{\pm}$ are the zero-zero
components of the metric tensor in the outer and inner regions of the
bubble; $G$ is the gravitational constant, $R$ is the bubble radius
and the overdot denotes the derivative with respect to proper time
$\tau$ on the shell.

Let  us  investigate a general case of a spherically symmetrical charged
bubble, when the metric outside the bubble is
$$
f_+ = 1 - \Delta - \frac{2Gm}{r} + \frac{Ge^2}{r^2} - G\Lambda_+r^2 ,
$$
while inside we have
$$
f_- = 1 - G\Lambda_-r^2 ,
$$
where $\Lambda_\pm \equiv \frac{8\pi}{3}\cdot\rho_\pm$ , $\rho_\pm$
being the vacuum energy density in the outer and inner regions. The
parameters $m$ and $e$ are the Schwarzschild mass and the charge of
the shell, respectively, and $g^{\infty}_{00} = 1 - \Delta$ is
the value of the metric tensor in the infinity.

Now the equation of motion(\ref{5}) takes the form
$$\sqrt{\dot R^2 + 1 - \Delta - \Lambda_+GR^2 - \frac{2Gm}{R} +
\frac{Ge^2}{R}} - \sqrt{\dot R^2 + 1 - \Lambda_-GR^2} = -\kappa GR
.$$
Finding $m$ from this equation we obtain:
\begin{equation}
m = - \frac{\Delta}{2G}\cdot R - \frac{a}{2}\cdot R^3 +
\frac{e^2}{2R} + \kappa R^2\cdot \sqrt{\dot R^2 + 1 - \Lambda_-GR^2}
, \label{6} \end{equation}
where $a \equiv \Lambda_+ - \Lambda_- + G\kappa^2$.

It is easy to understand the meaning of terms in (\ref{6}). The first
term is the asymptotical energy of the Higgs field forming the
bubble. The second term represents the volume energy of the bubble (a
difference between the old and new vacuum energy densities) and the
energy of gravitational self-interaction of  the  shell (the
surface-surface binding energy). The third term is the electrostatic
energy lying in the three-space outside the bubble. The last term
contains the kinetic energy of the shell and the surface-volume
binding energy.

Introducing new dimensionless variables
\begin{equation}
z \equiv \frac{Rb^{1/6}}{(-2m)^{1/3}} ,
~~~~~\tau' \equiv \frac{\tau b^{1/2}}{2\kappa} , \label{7}
\end{equation}
where $b = a^2 + 4\kappa^2\Lambda_- G,$
and dimensionless parameters
$$ A \equiv ab^{-1/2} ,~~~~~ E \equiv -
4\kappa^2(-2m)^{-2/3}b^{-2/3} ,$$
$$ Q^2 \equiv e^2(-2m)^{-4/3}b^{1/6} ,~~~~~ D \equiv \Delta
(-2m)^{-2/3}b^{1/3} , $$
we can represent the equation of motion (\ref{6}) as
$$ \biggl( \frac{dz}{d\tau'} \biggr)^2 + U(z) = E , $$
which is identical to that of the point-like particle  with
the energy $E$, moving in one dimension under the influence of the
potential
\begin{equation} U(z) = - \Biggl[ z^2 -
\frac{2A}{z}\cdot\biggl(1+\frac{Q^2}{z}+ Dz \biggr) +
\frac{1}{z^4}\cdot\biggl(1+\frac{Q^2}{z} + Dz \biggr)^2 \Biggr] ,
\label{8}
\end{equation}

In the equilibrium state
$$
\dot z_{|z=z_0} = 0 ,~~~~~\biggl(\frac{\partial
U(z)}{\partial z}\biggr) _{z=z_0} = 0 , $$ where $z_0$ is the
equilibrium point,  $U(z_0) = E$ and one can find the critical mass
and the equilibrium radius of the bubble
\begin{equation} m_0 = - \frac{4\kappa^3}{b{U_0}^{3/2}} ,~~~~~
 R_0 = \frac{2\kappa z_0}{b^{1/2}{U_0}^{1/2}} ,
 \label{9}
 \end{equation}
where $U_0 = |U(z_0)| > 0$. Note that $m_0$ is negative for the
positive $b$.

For the real trajectories potential (\ref{8}) must be negative since
$E < 0$.  Such a potential, for the case of uncharged shells, $Q =
0$, and with $ m > 0$, was discussed in \cite{BGG87,AKMS87}, while
for the case of Minkowskian asymptotics, $D = 0$ and $m < 0$, in
\cite{BG94}. Investigating potential (\ref{8}) in \cite{BG94} we have
found that in case when $D = 0$ it has the single maximum and
equilibrium state with (\ref{9}) is unstable for any values of
parameters. However in the case when $D \neq 0$ the term $Dz$ in
(\ref{8}) for some values of $D$ causes the appearing of a minimum of
the potential and gives the stable configuration.

Here we would like to note that sometimes for applications it is
more easy to evaluate the critical radius and mass of the bubble
directly from the equation (\ref{6}) imposing the conditions
\cite{BKT87}
\begin{equation} \dot R = 0 , \quad
\left. \frac{\partial m(R,\dot R)}{\partial R}\right. |_{\dot R=0} =
0 .
\label{10}
\end{equation}

The sign of the last term in equation (\ref{6}) is principal when we
investigate the problem of stability of the spherical shells. For the
ordinary matter this sign is negative, thus
$$ \left.  \frac{\partial
m(R,\dot R)}{\partial (\dot R^2)}\right. |_{\dot R=0} < 0 $$
and the equilibrium state (\ref{10}) is stable if the function
$m(R,\dot R=0)$ takes a maximum value at the point $R_0$
\cite{BKT87}.  For the case of domain walls, due to Tolman's formula
(\ref{2}), sign of the last term in (\ref{6}) is positive,
 $$ \left.\frac{\partial m(R,\dot R)}{\partial (\dot R^2)}\right.
 |_{\dot R=0} > 0 ,$$
and the equilibrium state (\ref{10}) is stable if the function
$m(R,\dot R=0)$ takes a minimum at the point $R_0$ \cite{BKT87}.

Now let us discuss some particular cases.

The simplest example of the antigraviting stable configuration is the
case of the Minkowski metric inside the bubble, $f_- = 1$, and the
Schwarzschild metric with the non-Minkowskian asymptotics, $f_+ =
1 - \Delta - 2m/r$, outside the bubble. In this case equation
(\ref{6}) has the form:
\begin{equation} m = -
\frac{\Delta}{2G}\cdot R - \frac{G\kappa^2}{2}\cdot R^3 + \kappa
R^2\cdot \sqrt{\dot R^2 + 1} .  \label{10a} \end{equation} From this
equation it is easy to find using (\ref{10}) the radiuses of
the critical configurations:  $$ R_0 = \frac{2 \pm \sqrt{4-3\Delta
}}{3G\kappa }, $$ one of which (with the lower sign) is stable, since
$\partial^2 m/\partial R^2 |_{\dot R =0} > 0$ for this value of $R$.

Inserting the value of $R_0$ into (\ref{10a}) one can find that the
mass of such configuration is negative.

In more simplified case, if we neglect the second term in equation
(\ref{10a}), the critical radius and mass of the configuration are
$$
R_{0} = \frac{\Delta}{4G\kappa}, \quad m_0 =
-\biggl(\frac{\Delta}{4G\kappa}\biggr) ^2 .  $$
This is a stable configuration, since $R_0$ is a minimum point
of the function $m(R, \dot R =0)$.

As the other example let us consider the case when the surface
density $\sigma$ in equation (\ref{6}) can be neglected. However,
as it was mentioned above, its sign governs the stability of the
system. From  relations (\ref{10}) for this case one can find
\begin{eqnarray}
\label{100}
R^{2}_{0} =
\frac{\Delta}{6G(-a)}\cdot\biggl( 1 + \sqrt{1 -
\frac{12ae^2}{\Delta^2}}\biggr) , \\
m_0 =  \frac{\Delta
R}{G}\cdot\biggl( -2 + \sqrt{1 - \frac{12ae^2}{\Delta^2}}\biggr) .
\label{101}
\end{eqnarray}
From this relations one can notice that the stable configuration is
possible only for the negative $a = \Lambda_+ - \Lambda_-$. The sign
of the mass of the critical bubble depends on values of parameters
$a, e$ and $\Delta$ and for different models can be positive,
negative or zero.

The next example of the stable spherical remnant of the false vacuum
surrounded by a spherical domain wall and with non-Minkowskian
asymptotics is the global monopole.  We want to treat the monopole
problem within the thin-wall approximation, i.e. we could regard that
the whole variation of the scalar field forming the monopole is
concentrated near some value of the radius $R_{0}$. In the spherical
coordinates the zero-zero component of the energy-momentum tensor of
the global monopole configuration reads (see for example
\cite{HL90}):
\begin{equation}
T^0_0 = \eta^2\biggl[\frac{\xi^2}{r^2}
+ \frac{1}{2}\biggl(\frac{\partial \xi}{\partial r}\biggr) ^2 +
\frac{\lambda}{4}\eta^2(\xi^2 - 1)^2\biggr] ,
\label{11}
\end{equation}
where $$ \xi = 0~~~~~  r < R_0 ,$$ $$ \xi = 1~~~~~  r > R_0 . $$ In
other words, we are modeling the monopole by a pure false vacuum
inside the core, and an exactly true vacuum at the exterior. In the
outer region (\ref{11}) does not contains a constant term. Thus for
the monopole the outer vacuum energy density $\rho_+$ is zero.  In
the inner region $\rho_- = \lambda \eta^4/4$, and we have
\begin{equation} e = \Lambda_+ = 0,~~~~~ \Lambda_- =
\frac{8\pi}{3}\cdot \frac{\lambda \eta^4}{4} .  \label{12}
\end{equation}

For the surface density $\sigma$ of the monopole within the thin-wall
approximation we find
\begin{equation} \label{13} \sigma = \int^{R_0
+ \delta}_{R_0 - \delta} T^{0}_{0} dr = \int^{R_0}_{R_0 - \delta}
 \frac{\lambda\eta^4}{4} dr + \int^{R_0 + \delta}_{R_0}
 \frac{\eta^2}{r^2} dr \approx \biggl(\frac{\lambda\eta^4}{4} +
 \frac{\eta^2}{R_0^2}\biggr)\delta , \end{equation} where $\delta \ll
R_0$ is the width of the wall. Thus  $\kappa \equiv
4\pi\sigma \ll \Lambda$ and we can neglect it in  equation
(\ref{6}) for the monopole.

We can find the quantity $\Delta$ in (\ref{6}), formed from the first
term of (\ref{11}), from the solution of Einstein's equations for the
monopole:
\begin{equation} \label{14} g_{00} = 1 + \frac{8\pi
G}{r}\int^{\infty}_{0}T^{0}_{0}r^2dr = 1 - \Delta - \frac{2Gm}{r}
,\end{equation} where \begin{equation} \label{15} \Delta = \frac{8\pi
G}{r}\int^{r}_{\infty}T^{0}_{0}r^2dr = 8\pi G\eta^2 \neq 0 .
\end{equation}

Using expressions (\ref{12}), (\ref{13}) and (\ref{15}) from
(\ref{100}) and (\ref{101}) for the monopoles radius and mass we
find:
$$ R_0 \sim \bigl(\frac{\Delta}{3G\Lambda_-}\bigl)^{1/2} =
 \frac{2}{\eta\sqrt{\lambda}} , $$ $$ m \sim -
 \frac{\Delta^{3/2}}{3G^{3/2}\Lambda^{1/2}_{-}} = -
 \frac{16\pi\eta}{3\sqrt{\lambda}} < 0 .$$
These values are in good  agreement with the exact solutions for the
global monopole obtained in paper \cite{HL90}.

At the end we would like to notice that in case of t'Hooft-Polyakov's
monopole the gauge field energy cancels nonzero energy of scalar field
in infinity. Thus $\Delta = 0$ and we have the ordinary Schwarzschild
metric, as it was considered in paper \cite{CF75}.\vskip 0.5cm

The research described in this publication was made possible in part by
Grant MXL200 of Georgia Government and the International Science
Foundation.

\newpage
 
\end{document}